# Data load Balancing in Mobile ad hoc network using Fuzzy logic (DBMF)

Abu Sufian[*1], Farhana Sultana[1], Prof. Paramartha Dutta[2]

[*1]Department of Computer Science, University of Gour Banga, Malda, West Bengal, India
[*]sufian.csa@gmail.com
[2]Department of Computer & System Sciences, Visvha-Bharati University, Santiniketan, West Bengal, India

**ABSTRACT**

Volume and movement of data rapidly increasing in every type of data communications and networking, and ad hoc networks are not spare from these challenges. Traditional Multipath routing protocols in Mobile Ad-hoc Networks (MANETs) did not focuses data load distribution and balancing as much as required. In this scheme, we have proposed data load distribution and balancing through multiple paths simultaneously. We have considered three important parameters of ad hoc network those are: mobility of node, energy of node and packet drop rate at a node. This scheme combines these three metrics using fuzzy logic to get decisive parameter. We have shown improvement of this scheme over similar kind of protocols in NS-2 network simulator.

## I. INTRODUCTION

Mobile Ad-hoc Networks (MANETs) is a self-organized, infrastructure less network, where a group of mobile nodes which are capable of sending and receiving radio signals can quickly make this type of networks, where a node can play a role as a sender, as a receiver or as a router[1, 2]. Such kind of network is very useful in battlefield communication, communication after natural disaster or at the time of rescue operations and in many more situations where traditional network failed to deliver desired service on those situations [3]. Many useful routing protocols have been proposed in MANETs since mid-1990, some are table driven (proactive) such as DSDV[4], OLSR[5], some are on-demand(reactive) such as AODV[6], DSR[7], AOMDV[8] and some are hybrid( both proactive and reactive) such as ZRP[9], EMR-PL[10].

This type of network has two main challenges: firstly, nodes are dynamic in nature, so connection among nodes is temporary, secondly nodes are energy constrained, that is nodes are battery powered. Therefore, chances of connection breakage are very high and when connection breaks, then huge number of Route Request (RREQ) packets-need to be generated to setup new connection [11,12]. Volume and movement of data are increasing rapidly in MANETs compare to other kind of networks due to high resolution multimedia data and live streaming.

To resolve such kind of challenges in MANETs, the researchers have adopted several initiatives such as multipath routing [13], route switching[14], scheduling [15] But most of the proposed multipath routing protocols focused single track (path) communications at a time, and other path(s) are reserved to increase network stability and future use, so data transfer rate are not as fast as required. In order to tackle these challenges, we have proposed Data load Balancing in Mobile ad hoc network using Fuzzy logic (DBMF). This scheme considers three important parameters: mobility and residual energy of node along with packet drop rate to calculate multiple paths from source to destination based on fuzzy logic, and then distribute data simultaneously through selected multiple paths. Therefore, this scheme could deliver huge volumes of data from source to destination in short span of time. We have tested our scheme using NS-2 [16] networks simulator and compare with similar kind of schemes such as MMRE-AOMDV [17], ZD-AOMDV [18] etc.

Rest of the article is organized as follows: in section 2 literature review has been done, section 3 explains scheme details, section 4 describes simulation studies and conclusion is in section 5.

## II. LITERATURE REVIEW

Many load balancing schemes have been proposed from initial development time of mobile ad-hoc networks to modern day's sophisticated routing schemes. These load balancing strategies become very important as volume and movement of data is increasing rapidly day by day. In this section we briefly reviewed some of the most important load balancing scheme.

M.R. Pearlman et.al in [19] had proposed a load balancing scheme at very early stage of development of MANETs. This was one of the first routing scheme where data transfer delay was reduced about 20% compare to the standard routing schemes of that time. After that Linifang Zhang et.al proposed a load balancing scheme in [20] which multiple path are selected based on Multipath Source Routing(MSR) [21], and then the input traffic is distributed among these selected multiple paths. Y. Ganjali and A. Keshavarzian introduced load balancing scheme of wired network in MANETs after making some modifications [22]. In this scheme data traffic is evenly distributed into specified number (K) of shortest selected paths. Salman Ktari et.al proposed routing protocol called LOBAM [23] based on OLSR [5] where data packets are distributed among multiple paths in consideration of neighborhood load. Another load balancing scheme [24], proposed by Amir Darehshoorzadeh et.al where load is distributed simultaneously through all the selected multiple path so that energy consumption can be reduced. R. Vinod Kumar and R. S. D Wahida Bannu suggest Load balancing approach for AOMDV in Ad-hoc Networks [25]. They used a threshold to identify load of intermediate nodes, if load reaches to threshold point then this load is distributed among other paths. Yaha M. Tashtouch and Omar A. Darwish proposed a Load Balancing scheme based on Fibonacci number sequence, called FMLB [26]. Another load balancing scheme LBAOMDV [27] based on AOMDV [8] was proposed by Saleh A. Alghamdi. Here load is distributed into those paths which has maximum residua energy.
Anusuman Bhattacharya and Koushik Sinha proposed a routing protocols, called LCMK [28], this protocol first selects multiple paths from source to destination based on AODV [6], after that load is distributed among selected paths in such a way that the number of data packets sent through a path in such a manner that the number of data packet sent through this path is increasingly proportional to the routing time of that path. Yumei Liu et.al propose a multipath routing protocol focusing energy efficient and load balancing, called MMRE-AOMDV [17], where they extends the AOMDV [8] routing protocol. This scheme finds the minimal nodal residual energy of each path in the process of selecting multiple path then sorts these path based on nodal residual energy in decreasing order. This scheme increases life time of networks and balances the data traffic load through energy efficient path. Nastooh Taheri Javan et.al proposed a zone disjoint multipath scheme called ZD-AOMDV [18] based on AODV [6]. This scheme first selects multiple paths from source node to destination node through different zone, and then simultaneously delivers data packets through these paths. This scheme balances load and reduces end to end delay.

## III. DETAILS OF DBMF
### A. Overview

In DBMF, every node broadcasts 'Hello' message to its all neighbors within its radio ranges. All the downlink neighbor nodes receive that 'Hello' message and reply back by 'Ack' message to maintain their links. This scheme assumes that, the link can be broken in two situations: firstly, nodes can go away from radio ranges and that is the cause of movement or mobility of nodes and secondly participating nodes are running out of its energy (battery power of node). Considering these two situations, DBMF calculate link duration based on these two parameters which are link duration prediction based on mobility of node (LPM) and Link duration prediction based on energy of node (LPE). And then calculate packet drop rate of that link.

In link duration based on mobility of node, the relative mobility of a node with respect to its uplink node can be calculated based on received signal strength of specified number of previous 'Hello' messages sent by the uplink node. This link duration can be found by the ratio of current distance with relative mobility. On the other hand link duration based on energy of node can be calculated as: data load arrival rate and data load departure rate along with additional load (Which can be assigned if other link is broken or new communication is started by other nodes) by this it can be estimated that how long a node can be operable (a node can be operable if it has at least 40% energy left [29]). Then combining these two different weights with drop rate, we can calculate efficiency of paths, after that data load can be distributed among selected paths according to their combining weight. These calculation is done based on fuzzy logic, because it is simple, can tolerate imprecise data and has soft computing capabilities [30].

### B. Link duration prediction based on mobility of node (LPM):

Suppose a node $n_j$ belonging within radio-frequency range of node $n_k$ and received last n number of 'Hello' messages succesfully. Received signal strength of i-th 'Hello' message at node $n_j$ is 'rec_pow_sig$_k$(j, i)' and when it sent by the node $n_k$ is 'trans_pow_sig(k)'. Let distance between node $n_j$ and node $n_k$ at the time of i-th 'Hello' message is 'distance$_i$(j, k)'. Therefore, by Frii' transmission equation of antenna theory, we can write:

$$rec\_pow\_sig_k = \frac{K * trans\_pow\_sig(j)}{distance_j(j,k)^q}$$

or, $$distance_i(r,s) = \sqrt{\frac{K * trans\_pow\_sig(j)}{rec\_pow\_sig_k(j,i)}} \quad (1)$$

where K is constant and values of q depends on medium and it could be 2 or 3.

If 'intv' is the time interval between two successive 'Hello' message and 'rad_rng(j)' is radio range of node $n_j$, then average relative mobility of node $n_k$ is 'avg_rel_mob(k, j)' with respect to node $n_j$ can be found as:

$$Avg\_rel\_mob(k,j) = \sum_{i=1}^{n} \frac{distance_{i+1}(j,k) - distance_i(j,k)}{n * intv} \quad (2)$$

The link between node $n_j$ and $n_k$ can be broken if node $n_j$ go away from radio range of node $n_k$ and this can be predicted based on average effective relative mobility calculated in equation(2) by the following equation(3):

$$LP(j,k) = \frac{rad\_rng(j) - distance(j,k)}{Avg\_rel\_mob(j,k)} \quad (3)$$

This link prediction LP is positive unbounded variable and unbounded variable is difficult to modeling. Therefore, logistic function is used to make it bounded from 0 to1, which is a fuzzy variable as below:

$$LPM(j,k) = 2x \frac{1}{1 + e^{LP(j,k)}} - 1$$

or, $$LPM(j,k) = \frac{1 - e^{LP(j,k)}}{1 + e^{LP(j,k)}} \quad (4)$$

Equation (4) shows a value of LPM is within 0 to 1. Value near 0 means worst and 1 means best.

### C. Link duration prediction based on energy of node (LPE)

In this scheme, a node can participate more than one paths as this scheme is link dis-joint. Let a node $n_j$ participate 'np' number of paths and data packet arrival and departure rates are 'data_arr$_j$' and 'data_dept$_j$'. Maximum data packet arrival and departure capacities are 'Max_data_arr$_j$' and 'Max_data_dept$_j$' respectively in per unit of time. It is also assumed that initial battery power is 'Eng$_j$' and current power is 'C_Eng$_j$' of node $n_j$. it is already mentioned that at least 40% of total battery power is required to remain operational. If the node $n_j$ spent 'pow_req_recv(j)' and 'pow_req_dept(j) unit of energy to receive and forward each data packet then total energy spent per unit time would be: {$data\_arr_j * pow\_req\_recv(j) + data\_dept_j * pow\_req\_dept(j))$}

As we already mentioned node $n_j$ can participate np number of paths, so after assigned current request, another np-1 path's data transfer request can be arrived at node $n_j$. Then new data packet arrival rate will be

$$data\_arr(j) = data\_arr_j + \sum_{i=1}^{np-1} data\_arr_i$$

or, $$data\_arr(j) = \sum_{i=1}^{np} data\_arr_i \quad (5)$$

The revised data departure rate would be

$$data\_dept(j) = data\_dept_j + \sum_{i=1}^{s} data\_dept_i \quad (6)$$

The above rate must satisfy following inequalities:
 $data\_arr(j) \leq Max\_data\_arr_j$ and
 $data\_dept(j) \leq Max\_data\_dept_j$

Therefore, the node $n_j$ spent energy per unit for transmission is calculated as:
 $Trans\_eng(j) = pow\_req\_dept(j) * data\_dept(j)$
or,
$$Trans\_eng(j) = pow\_req\_dept(j) * (data\_dept_j + \sum_{i=1}^{s} data\_dept_i) \quad (7)$$

and node $n_j$ spent maximum energy per unit of transmission is calculated as:
$rec\_eng(j) = pow\_req\_recv(j) * Max\_data\_arr_j$ (8)

Therefore, by equation (7) and (8), we can calculate total energy spent by the node $n_j$ is *(trans_energy(j) + recv_eng(j))*.

Let 'Total_No_Packet' be the total number of data packets to be transfer and $T_j$ be the time required for each packet by the node $n_j$. So, till node $n_j$ need to be active is given by equation (10):

$$Active\_time(j) = \frac{Total\_No\_Packet}{T_j} \quad (5)$$

Therefore, the node $n_j$ reduces it battery power for entire multipath communication and it can be expressed in equation (10) as:

$$Total\_energy\_cons(j) = Active\_time(j)*(trans\_energy(j)+recv\_eng(j)) \quad (10)$$

Now the node $n_j$ would be remain operational if the following inequality holds:

$Eng_j – C\_eng_j – Total\_energy\_cons(j) \geq 0.4\ Eng_j$
or $0.6Eng_j – C\_eng_j – Total\_energy\_cons(j) \geq 0$

Therefore, link duration prediction based on energy of node can be expressed for link (j, k) as:

$$LE(j,k) = \begin{matrix} Min\{\ Active\_time(j), Active\_time(k)\} \\ 0 \end{matrix} \quad (11)$$

**NB:** If above mentioned inequality holds for both nodes nj and nk, then *LE(j,k)* will take value *Min{ Active_time(j), Active_time(k)}* otherwise will take 0. Here, LE also a positive unbounded variable, and is also required to map into a fuzzy variable LPE, which can be done by same logistic function as:

$$LPM(j,k) = \frac{1 - e^{LE(j,k)}}{1 + e^{LE(j,k)}} \quad (12)$$

By the equation (12), the value of LPM could be from 0 to 1 and here 0 means worst and 1 means best.

**D. Data packet drop rate calculation**

Data packet drop rate count can be easily calculated by subtracting number of data packets departed from number of data packet arrived at a node. Therefore, number of data packet droppped at node $n_j$, that is 'Data_drop$_j$' can be calculated as:

$$Data\_drop_j = data\_arr(j) – data\_dept(j) \quad (13)$$

Therefore reverse data packet ratio 'DPR' at the node $n_j$ can be calculated as:

$$DPR = \frac{1 - Data\_drop_j}{data\_arr(j)} \quad (64)$$

Clearly this 'DPR' is a fuzzy variable. Its value is zero if all packets dropped means worst-case and value is 1 means no drop.

**E. Overall Path duration prediction**

These two LPM and LPE calculated in equation (4) and (12), range between 0 and 1. LPM and DPR are uniformly divided into four crisp ranges such as: 0-0.25 is indicated as fuzzy premise variable a, 0.25-0.50 as b, 0.50-0.75 as c and 0.75-1.00 as d, and LPE is divided into four crisp ranges such as: 0-0.40 as a, 0.40-0.60 as b, 0.60-0.80 as c and 0.80-1.00 as d. By combining both LPM and LPE with equal priorities by fuzzy rule bases produce temporary variable 'TM' in table-1 then overall weight called 'Link_Life(j, k)' between node $n_j$ and $n_k$ can be calculated by combining this 'TM' with 'DPR' in table-2. Link_Life(j,k) is also uniformly distributed into four crisp ranges a, b, c and d same as LPM.

TABLE 1.

FUZZY COMBINATION OF LPM AND LPE TO PRODUCE TEMPORARY VARIABLE TM.

| LPM ➡ <br> LPE ⬇ | a (0-0.25) | b(0.25-0.50) | c(0.50-0.75) | d(0.75-1.00) |
|---|---|---|---|---|
| a(0-0.40) | a | a | a | a |
| b(0.40-0.60) | a | b | c | c |
| c(0.60-0.80) | b | c | c | d |
| d(0.80-1.00) | c | c | d | d |

TABLE 2.

FUZZY COMBINATION OF TM AND DPR TO PRODUCE REQUIRED PARAMETER LINK_LIFE .

| TM ➡ <br> DPR ⬇ | a(0-0.25) | b(0.25-0.50) | c(0.50-0.75) | d(0.75-1.00) |
|---|---|---|---|---|
| a(0-0.25) | a | a | a | a |
| b(0.25-0.50) | a | b | c | c |
| c(0.50-0.75) | b | c | c | d |
| d(0.75-1.00) | c | c | d | d |

Therefore, overall route life time 'Route_Life(R)' can be calculated by the following equation (14). Here R is a route through nodes $n_1, n_2, n_3, …, n_{h-1}, n_h$. By the equation (14) we can say that if any link breaks, route will break. This scheme will give ranks based on equation (14) to all possible paths from source to destination such as: $R_1, R_2, R_3, ..., R_{np}$. Among np number of ranked paths (route) s number are selected for simultaneous data transfer.

$Route\_Life(R) = Min\{ Link\_Life(1,2), Link\_Life(2, 3), Link\_Life(3,4), ..., Link\_Life(h-1, h)\}$ (14)

### F. Data Load distribution through multiple paths

DBMF gives 'Rout_Life' value to every possible path in order to select multiple paths from source to destination, and as mentioned in previous sub-section that s number of paths are selected to distribute the data packets to balance the load among all selected paths. In this scheme data packet distribution is done in such a way that data load is balanced as well as all assigned path completed transferring assigned portion of data packet almost at the same time with minimum data packet drop and overall in less transaction time.

Suppose i-th path among s number of selected paths, carry '$pckt\_prtn_i$' and each packet take '$delay_i$' unit time to reach destination where $1 \leq i \leq s$. Therefore, the ideal condition to finish almost same time as follows:

$pckt\_prtn_1 * delay_1 = pckt\_prtn_2 * delay_2 = pckt\_prtn_3 * delay_3 = ... = pckt\_prtn_s * delay_s = PD$ (say)

i.e,
$$pckt\_prtn_i = \frac{PD}{delay_i}$$ (15)

Where,
$$Total\_No\_Packet = \sum_{i=1}^{s} pckt\_prtn_i$$

by equation(15), it is clear that each selected path will get a portion of total data packets which is inversely proportional to the packet travelling time of that path.

## IV. SIMULATIONS

This scheme adopts AODV [8] as base routing protocol and simulation is carried out in NS-2 [17]. Our results compare with two load balancing scheme namely MMRE-AODV[18] which is single track and ZD-AOMDV[19] which is multiple track load balancing scheme. The minimum system requirements mentioned in table-3.

TABLE 3.

THE MINIMUM SYSTEM REQUIREMENTS FOR SIMULATION THIS SCHEME.

| Components | Specification |
| --- | --- |
| Processor | 800 MHz Pentium IV |
| RAM | 256 MB |
| Hard Disk | 40 GB |
| Operating System | Red Hat Linux 6.2 |
| Network Size(No. of Nodes) | 20, 50, 100, 150 and 200 |
| Network Area | 500 meter by 500 meter |
| Speed of nodes | 5 m/s, 10 m/s, 25 m/s and 50 m/s |
| Radio Ranges | 10 meter to 50 meter |
| Mobility model | Random Waypoint [9] |

### A. Experimental Results

We have considered three metrics; first one is packet delivery ratio, which is percentage of data packet successfully delivered to indeed destination, second one is average delay, which is the time required to send a data packet from source to destination in milliseconds, and third one is packet drop rate, which is number of data packets drop per second by a node. Different output of these three metrics is taken against different network sizes namely 20 nodes, 50 nodes, 100 nodes, 150 nodes and 200 nodes.

Packet delivery ratio versus Size of networks for all the three algorithms shown in Fig. 1, Average delay versus Size of networks shown in Fig. 2 and Packet drop ratio versus Size of network shown in Fig. 3. In Fig.1 we can see that DBMF-AODV in NS-2 simulator, uniformly outperform over MMRE-AODV and ZD-AOMDV in all five types of network size which we have consider. In Fig. 2 it is clear that DBMF-AODV gives less delay than other two and performs better in respect to increasing network size. Packet drop rate also less in the proposed scheme compare to other two. It is clear from Fig. 3, packet drop rate increases sharply for MMRE-AODV and ZD-AOMDV on increasing network size, whereas for DBMF-AODV marginally.

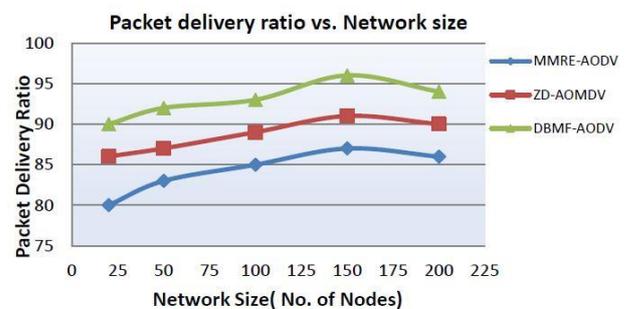

**Figure 1**. Packet delivery ratio vs. Network size.

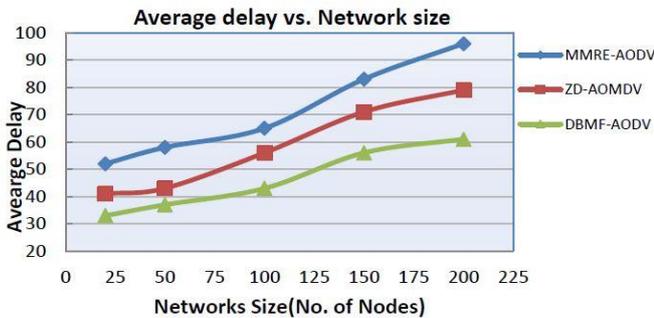

**Figure 2**. Average delay vs. Network size.

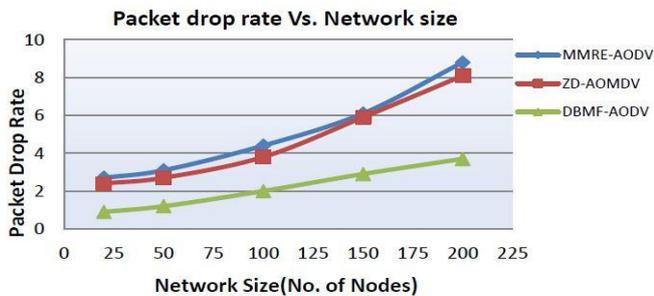

**Figure 3**. Packet drop rate vs. Network size.

## V. CONCLUSION

As capability of every IoT equipment (node in MANETs) increases with technological improvements, therefore volume of data is also increasing rapidly. To handle such challenges data packet load balancing scheme can be adopted by embedding this with existing routing protocols in MANETs to improve performance, and this DBMF is such kind of scheme. Three main contribution of DBMF are: DBMF increases networks life time, it reduces average data packet delivery delay and reduces data packet drops.